\documentclass[aps,pra,amsmath,amsfonts,amssymb,twocolumn,nofootinbib]{revtex4}

\usepackage{amssymb}
\usepackage{graphicx,psfrag,color}
\usepackage{dcolumn}
\usepackage{bm}
\usepackage{mathbbol}
\usepackage[T1]{fontenc}

\newcommand{\med}[1]{\left\langle\! #1 \!\right\rangle}

\begin{document}
\flushbottom

\title{Detecting quantum critical points at finite temperature via quantum
teleportation}

\author{G. A. P. Ribeiro}
\author{Gustavo Rigolin}
\email{rigolin@ufscar.br}
\affiliation{Departamento de F\'isica, Universidade Federal de
S\~ao Carlos, 13565-905, S\~ao Carlos, SP, Brazil}

\date{\today}

\begin{abstract}
We show that the quantum teleportation protocol is a powerful tool to
study quantum phase transitions (QPTs) at finite temperatures. We consider 
a pair of spins from an infinite spin-1/2 chain (XXZ model) in equilibrium with a reservoir at temperature $T$ as the resource used by Alice and Bob to implement
the teleportation protocol. We show that the efficiency of this pair of spins to
teleport a qubit is drastically affected after we cross a quantum critical point
(QCP), even for high values of $T$. Also, we show that the present tool
is as sharp as quantum discord (QD) to spotlight a QCP, 
where QD is the best finite $T$ QCP detector known to date. Contrary to QD, however,
we show that the present tool is easier to compute theoretically and has a direct experimental and operational meaning. 
\end{abstract}


\maketitle

\section{Introduction} 

A QPT is the abrupt change
in the physical properties of a many-body system that occurs at
the absolute zero temperature ($T=0$) while we change its Hamiltonian $H$ \cite{sac99}.
At $T=0$ the 
system is completely described by its ground state, 
a function of $H$. If we slowly modify $H$
we may reach a QCP in the parameter space
where the macroscopic properties of the system 
abruptly change. This ``change of phase'' is driven 
solely by quantum fluctuations and is 
almost always characterized by a fundamental symmetry change in the system's ground state and the emergence of an order parameter. For a magnetic system, for instance,
the order parameter is the net magnetization which becomes non-null after the system
enters the ferromagnetic phase.

The physical principle behind the quantum fluctuations is the Heisenberg 
uncertainty principle. There is no role for thermal fluctuations in a QPT
since we are in principle at $T=0$ and to drive the system from one phase to another we 
usually change a single quantity (tuning parameter) of $H$ . Examples of 
tuning parameters are the coupling 
constants or external fields acting on the system.   
The superfluid-Mott insulator transition \cite{gre02}, the superconductor-insulator transition
\cite{gan10}, and the ferromagnetic-paramagnetic transition in
some metals \cite{row10} are paradigmatic examples of QPTs.

Many useful theoretical tools to investigate QPTs assume that we are exactly 
at $T=0$. For a spin chain the QCPs are obtained by studying the 
behavior of its magnetization,
bipartite \cite{wu04} and multipartite \cite{oli06}
entanglement, and more general quantum correlations \cite{dil08,sar08} 
as functions of the tuning parameter. 
The extremal values of these quantities or discontinuities in their first and second order derivatives are important indicators of a QPT. However, from the experimental point of view, the $T=0$ condition is unattainable due to the third law of thermodynamics and any small deviation from $T=0$ brings to the table thermal fluctuations that excite the system beyond the ground state,
limiting severely our ability to properly detect a genuine QCP. A remarkable 
example of this limitation is the inability of the entanglement 
of formation (EoF) \cite{woo98}
and the magnetic susceptibility to detect a QPT in spin chains
for $T>0$. For very small values of $T$, the former is already
zero before and after the QCPs and the latter is a smooth function of
the tuning parameter with no indication of QPTs when we cross the 
QCPs \cite{wer10,wer10b}.

In Ref. \cite{wer10b}, however, we showed that in the 
thermodynamic limit (infinite chains) thermal quantum discord (TQD) \cite{footnote1}
is a key theoretical tool that bridges the gap between the $T=0$ predictions of QPTs and the finite $T$ experiments. 
For several classes of spin chains, we showed that 
TQD detects QCPs for relatively high values of $T$ while at this same $T$ 
EoF is already zero and other thermodynamic quantities, including the magnetic susceptibility, are not able to correctly identify the QCP or are less efficient than TQD to properly identify it. 

Notwithstanding its tremendous success to detect a QCP at finite $T$
\cite{wer10b}, the QD \cite{oll01,hen01} has two handicaps.
First, the optimization problem that one needs to solve to obtain the QD 
is NP-complete \cite{hua14}. This means that the computation of 
QD is an intractable problem as we increase the size of the Hilbert space of the 
system under investigation. It is thus very difficult to
extend the analysis of spin-1/2 chains to higher spin chains \cite{mal16}. Second, 
QD does not have an operational meaning. 
There is no direct experimental procedure whose outcome is the QD. We need first
to obtain the system's density matrix and then use it to compute the QD. 

In this manuscript we present a tool that possesses all the outstanding
features of TQD in detecting QCPs at finite $T$ and, 
on top of that, does not have its two 
aforementioned handicaps. The present tool is based on the quantum teleportation 
protocol \cite{ben93}, where a pair of qubits in a spin chain is employed as the quantum resource
needed to implement the quantum teleportation protocol \cite{yeo02,rig17}. In Ref. \cite{rig17}, for two-spin systems
in equilibrium with a thermal reservoir at temperature $T$,
it was shown that the fidelity of the teleported state changes abruptly when we 
cross a QCP. Here we extend the results of Ref. \cite{rig17} to more general
settings, to more QCPs, and most importantly, we work in the thermodynamic limit. 

\section{The XXZ model} 

The spin-1/2 chain we study here in the thermodynamic
limit ($L\rightarrow \infty$) is given by the following Hamiltonian ($\hbar=1$),
\begin{equation}
H = \sum_{j=1}^{L}\left( \sigma^{x}_{j}\sigma^{x}_{j+1} +
\sigma^{y}_{j}\sigma^{y}_{j+1} + \Delta
\sigma^{z}_{j}\sigma^{z}_{j+1} \right). \label{Hxxz}
\end{equation}
We employ periodic boundary conditions and $\sigma_j^{z},\sigma_j^{y}, \sigma_j^{x}$
are the standard Pauli matrices associated with the qubit $j$. The anisotropy  
$\Delta$ is our tuning parameter and at $T=0$ this model has two
QCPs \cite{tak99}. At $\Delta = -1$ we have a first-order transition. The ground
state changes from a ferromagnetic phase ($\Delta < -1$) to the critical 
antiferromagnetic one ($-1 < \Delta < 1$). At $\Delta = 1$ we have a
continuous phase transition with the system becoming an Ising-like antiferromagnet
when $\Delta > 1$.

In thermal equilibrium with a reservoir at temperature $T$,
the density matrix describing this chain is $\varrho=e^{-H/kT}/Z$, with $Z=Tr[e^{-H/kT}]$ the partition function and $k$ Boltzmann's constant.
Tracing out from $\varrho$ 
all but two nearest neighbors gives the two-spin state \cite{wer10b}
\begin{equation}
\rho_{j,j+1}  \!=\! \left(\hspace{-.15cm}
\begin{array}{cccc}
\frac{1+\med{\sigma_j^z\sigma_{j+1}^z}}{4} & 0 & 0 & 0\\
0 & \frac{1-\med{\sigma_j^z\sigma_{j+1}^z}}{4}  &
\frac{\med{\sigma_j^x\sigma_{j+1}^x}}{2} & 0 \\
0 & \frac{\med{\sigma_j^x\sigma_{j+1}^x}}{2} &
\frac{1-\med{\sigma_j^z\sigma_{j+1}^z}}{4} & 0 \\
0 & 0 & 0 &  \frac{1+\med{\sigma_j^z\sigma_{j+1}^z}}{4}\\
\end{array}
\hspace{-.15cm}\right)\!\!. \label{rhoAB}
\end{equation}
The computation of the two-point correlation functions 
$\med{\sigma_j^\alpha\sigma_{j+1}^\alpha}=
Tr[\sigma_j^\alpha\sigma_{j+1}^\alpha\, \varrho]$, with $\alpha=x,z$,
in the thermodynamic limit for arbitrary $T$ and $\Delta$ 
was done in Refs. \cite{klu92,bor05,boo08,tri10} 
and reviewed in Ref. \cite{wer10b} (see Appendix \ref{apA}). 

\section{The teleportation protocol} 

\begin{figure}[!ht]
\includegraphics[width=8cm]{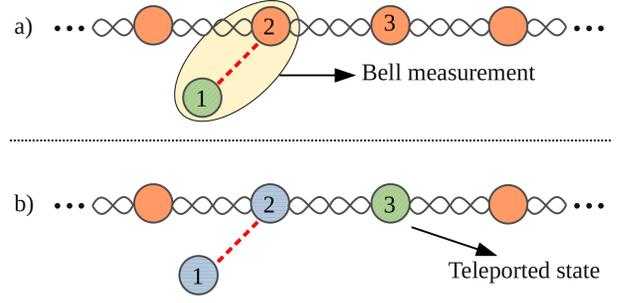}
\caption{\label{fig_scheme}(color online)  The teleportation 
protocol works as follows \cite{ben93}. 
Panel (a): First, one prepares the 
entangled resource (spins 2 and 3) to be used to teleport the input (spin 1).
Second, a Bell measurement (BM) is made by Alice on spins 1 and 2.
Panel (b): Third, Alice informs Bob of her BM result using a
classical communication channel. Fourth, Bob applies a 
unitary operation on spin 3, which depends on Alice's BM result,
to finish the protocol.}
\end{figure}

Equation (\ref{rhoAB}) describes the 
resource through which we implement the 
teleportation protocol. Setting $j=2$, 
the two-qubit state at sites $2$ and $3$
(see Fig.~\ref{fig_scheme}) is what one usually calls the state shared by 
Alice and Bob at the beginning of the teleportation protocol. The qubit 
to be teleported (spin $1$ in Fig.~\ref{fig_scheme})
is external to the chain
and can be prepared in any normalized pure state ($0\leq r \leq 1$ and $0\leq \gamma < 2\pi$),
\begin{equation}
|\psi\rangle
=r|0\rangle+\sqrt{1-r^2}e^{i\gamma}|1\rangle.
\label{step0}
\end{equation}

The initial state describing the three qubits before the beginning of the 
teleportation protocol is 
\begin{equation}
 \rho = \rho_{1} \otimes \rho_{23},
\label{stepA}
\end{equation}
where $\rho_1 = |\psi\rangle \langle \psi |$ and $\rho_{23}$ is given by 
Eq.~(\ref{rhoAB}). At the end of a given run of the protocol, i.e.,
after the four steps described in Fig. \ref{fig_scheme}, 
Bob's qubit (spin 3) is given by \cite{rig17}
\begin{equation}
\rho_{_{B_j}}=   \{U_jTr_{12}[P_j \rho P_j]U_j^\dagger\}/Q_j(|\psi\rangle),
\label{stepD}
\end{equation}
where $Tr_{12}$ is the partial trace on Alice's qubits (spins 1 and 2). 
In Eq.~(\ref{stepD}), $j$ denotes the BM result
obtained by Alice, namely, $j=\Psi^-,\Psi^+,\Phi^-,\Phi^+$, and $P_j$ 
the four projectors describing her BMs,
$P_{\Psi^{\pm}} = |\Psi^{\pm}\rangle \langle \Psi^{\pm}|$ and  
$P_{\Phi^{\pm}} = |\Phi^{\pm}\rangle \langle \Phi^{\pm}|$,
%
%
where the Bell states are
$|\Psi^{\mp}\rangle=(|01\rangle \mp |10\rangle)/\sqrt{2}$ and 
$|\Phi^{\mp}\rangle=(|00\rangle \mp |11\rangle)/\sqrt{2}$.
%

Alice's probability to measure Bell state $j$ is 
\begin{equation}
Q_j(|\psi\rangle) = Tr[{P_j \rho}]
\label{prob}
\end{equation}
and the unitary operation that Bob must implement on his qubit after being informed
of Alice's BM result is $U_j$. 

The unitary operation $U_j$ that Bob must implement on his qubit at the end of 
the protocol is also dependent on the entangled resource shared by Alice and Bob.
In its standard formulation \cite{ben93}, 
where they share a maximally entangled pure
state (Bell state), we have
\begin{equation}
S_{\Phi^+}=\{U_{\Phi^+},U_{\Phi^-},U_{\Psi^+},U_{\Psi^-}\}
=\{\mathbb{1},\sigma^z,\sigma^x,\sigma^z\sigma^x\}
\label{s1}
\end{equation}
if the shared Bell state is $|\Phi^+\rangle$, 
with $\mathbb{1}$ being the identity matrix,  and for 
$|\Phi^-\rangle$ and $|\Psi^\pm\rangle$, we have, respectively,
\begin{eqnarray}
S_{\Phi^-}=\{U_{\Phi^+},U_{\Phi^-},U_{\Psi^+},U_{\Psi^-}\}
=\{\sigma^z,\mathbb{1},\sigma^z\sigma^x,\sigma^x\}, \label{s2}\\
S_{\Psi^+}=\{U_{\Phi^+},U_{\Phi^-},U_{\Psi^+},U_{\Psi^-}\}=\{\sigma^x,\sigma^z\sigma^x,\mathbb{1},\sigma^z\}, \label{s3}\\  
S_{\Psi^-}=\{U_{\Phi^+},U_{\Phi^-},U_{\Psi^+},U_{\Psi^-}\}=\{\sigma^z\sigma^x,\sigma^x,\sigma^z,\mathbb{1}\}.\label{s4}
\end{eqnarray}
In the present case, 
the state $\rho_{23}$ shared by Alice and Bob is not pure and
is approximately described by one
Bell state in one phase and by a different one in another phase. Therefore, when characterizing the QCPs of a spin chain we will work with the four possible sets
of unitary operations above. As we will see, this approach is crucial to obtain the most efficient QCP detector based on the teleportation protocol.

To quantify the similarity between the teleported state, i. e., 
Bob's qubit at the end of the protocol (spin 3), with the input state teleported by 
Alice (spin 1), we use the fidelity \cite{uhl76}. For a pure input state we have
\begin{equation}
F_j(|\psi \rangle,S_k) = \langle \psi | \rho_{_{B_j}} | \psi \rangle,
\label{Fidj}
\end{equation}
where $| \psi \rangle$ is given by Eq.~(\ref{stepA}) and $\rho_{_{B_j}}$ by
Eq.~(\ref{stepD}). 
If the teleported state is exactly the input state, $F_j=1$,
and $F_j=0$ if the output is orthogonal to the input. Note that 
in general $F_j$ 
depends on the input state, the entangled resource shared
by Alice and Bob, and on the set of unitary corrections $S_k$ chosen by Bob. 
Here the entangled resource is fixed and given by Eq.~(\ref{rhoAB}), while we can
freely choose $|\psi \rangle$ and $S_k$, with $k=\Psi^\mp,\Phi^\mp$.

For a fixed input state, after several runs of the protocol the mean fidelity 
(efficiency) is \cite{gor06,rig15}
\begin{equation}
\overline{F}(|\psi\rangle,S_k)= \sum_{j=\Psi^{\mp},\Phi^{\mp}} 
Q_j(|\psi\rangle)F_j(| \psi \rangle,S_k). \label{Fbar}
\end{equation}
%
If we want an input state independent measure of the efficiency of the teleportation protocol, we can average over all states on the Bloch sphere. This 
is equivalent to considering in Eq.~(\ref{step0}) $r^2$ and $\gamma$ as independent
continuous random variables over their allowed values \cite{rig15,gor06}. 
Formally, this input state independent mean fidelity, average fidelity for short, can be written as
\cite{gor06,rig15,rig17,footnote1b}
\begin{equation}
\langle \overline{F}(S_k) \rangle = \int_\Omega \overline{F}(|\psi\rangle,S_k) 
\mathcal{P}(|\psi\rangle)d|\psi\rangle,
\label{Flangle}
\end{equation}
where we integrate over the sample space $\Omega$ comprised of all qubits on 
the Bloch sphere and $\mathcal{P}(|\psi\rangle)$ is the corresponding uniform probability distribution over $\Omega$. 

Before we specialize to the XXZ model, two remarks are in order.
First, from an 
experimental point of view, the present analysis is meaningful
when the time needed to execute the four steps of the teleportation protocol is
shorter than the time needed by the system to get
back to equilibrium with the thermal reservoir. The rate at which we
implement all the four steps of the protocol should be greater
than the thermal relaxation rate of the system. We should determine the state of the qubit teleported to Bob before it thermalizes
once again\footnote{The determination of the relaxation time is a tricky and non trivial problem, the calculation of which is beyond the scope of the present work. The relaxation time depends not only on the internal dynamics of the 
spin chain (its Hamiltonian) 
but also on how it interacts with the heat bath.}. 
Nevertheless, the experimental
procedure needed to teleport a qubit and measure Bob's state is clear and 
can in principle be implemented using 
state of the art techniques \cite{ron15,bra19,xie19,noi22,xue22,mad22,xie22}.
Knowing Bob's state at the end of the teleportation protocol 
for a representative sample of input states lying on the Bloch 
sphere is all we need to determine Eqs.~(\ref{Fbar}) and (\ref{Flangle}).
This should be contrasted with the determination of the EoF or the QD of a 
pair of spins. There is no direct experimental procedure to measure those quantities
for arbitrary mixed states. 
To compute those quantities we must have access to the complete 
density matrix describing the two qubits (spins 2 and 3) \cite{woo98,oll01,hen01}. 
For the XXZ model, for instance, we must have the complete knowledge of Eq.~(\ref{rhoAB}). On the other hand, 
to experimentally determine Bob's state at the end of the 
teleportation protocol we just need the single qubit density matrix describing it. 
We only need to measure one-point correlation functions instead of the two-point
ones needed for the computation of the EoF or the QD. 

Second, the choice for the fidelity to assess the similarity between 
Alice's input state and Bob's output state at the end of the teleportation
protocol is not mandatory. We could have used any other measure to quantify the similarity between those two states. As such, the present proposal to detect QCPs should not be confused with the ones based on the computation of the fidelity or the fidelity susceptibility between the system's whole ground state 
before and after the QCP \cite{qua06,zan06,buo07,lon07,gu08,gu08b}. 
Our proposal is conceptually different from the ones studied in the aforementioned references. In our proposal, the teleportation protocol plays an active and key
role in its implementation; no mention or use of the quantum teleportation protocol
are present in Refs. \cite{qua06,zan06,buo07,lon07,gu08,gu08b}. Also,
to fully implement the ideas of 
Refs. \cite{qua06,zan06,buo07,lon07,gu08,gu08b} without any approximation 
one needs to know the whole ground state of the chain before and after the QCP to compute the fidelity between those states. In our approach, from the theoretical point of view, we only need up to two-point correlation functions to compute the efficiency of the teleportation protocol. There is no need to know the whole ground state. Indeed, the determination of the whole ground state of the system requires much more information: one should in principle have access to all $n$-point correlation functions, where $n=1,2,3,\ldots, L$, with $L$ being the size of the spin chain. This clearly sets apart our approach (and the ones based on quantum discord and entanglement of formation between two spins) from those of Refs. \cite{qua06,zan06,buo07,lon07,gu08,gu08b}. Here we have a local approach,
i.e., we only need up to two-point correlation functions to implement our idea. In Refs. \cite{qua06,zan06,buo07,lon07,gu08,gu08b} we have a global approach,
i.e., we need the whole ground state to make progress. This is another feature that clearly sets apart both strategies. It also means that the present method allows
us to properly detect QCPs using local finite $T$ data alone. 

\section{Results} If we insert Eq.~(\ref{stepA}) into (\ref{prob}) we get 
$Q_j(|\psi\rangle) = 1/4$, for all $j$.
%
%
This means that Alice obtains with equal chances any one of the four 
Bell states after the BM on qubits 1 and 2. Note that 
the probabilities $Q_j$ are all independent of the input state (qubit 1). 
This is a particular feature of the XXZ model without external fields and 
can be traced back to the specific form of Eq.~(\ref{rhoAB}). 

Another feature of the present model is that
Eq.~(\ref{Fidj}) leads to
$F_{\Psi^\mp}(| \psi \rangle,S_k)=F_{\Phi^\mp}(| \psi \rangle,S_k)$.
%
%
In other words, $F_j$ is independent of $j$, i.e., independent of 
the outcome of Alice's BM. Therefore, for the present model,
Eq.~(\ref{Fbar}) can be written as
$\overline{F}(|\psi\rangle,S_k)= F_j(| \psi \rangle,S_k)$, for
any $j$.
%
%

If we maximize over all pure states and over $S_k$ we get for the overall maximum 
fidelity (see Appendix \ref{apB}),
%
%
\begin{equation}
\overline{\mathcal{F}}\hspace{-.02cm} =\hspace{-.1cm} \max_{\{|\psi\rangle,S_k\}}{\hspace{-.1cm}\overline{F}(|\psi\rangle,S_k)}
\hspace{-.07cm}=\hspace{-.07cm}
\max{\hspace{-.1cm}\left[\!\frac{1 \!+\! |\langle\sigma_2^z\sigma_{3}^z\rangle|}{2},\! 
\frac{1 \!+\! |\langle\sigma_2^x\sigma_{3}^x\rangle|}{2}\!\right]}\!.
\label{fmax}
\end{equation}
The extrema of Eq.~(\ref{Fbar})
occur for the input states $|\psi\rangle = |1\rangle$ and 
$|\psi\rangle = (|0\rangle + e^{i\gamma}|1\rangle)/\sqrt{2}$ (see Appendix 
\ref{apB}). 
Which state leads to the maximum (or minimum) fidelity depends on the phase 
of the spin chain and on the sign of the two-point correlation functions. 

\begin{figure}[!ht]
\includegraphics[width=8cm]{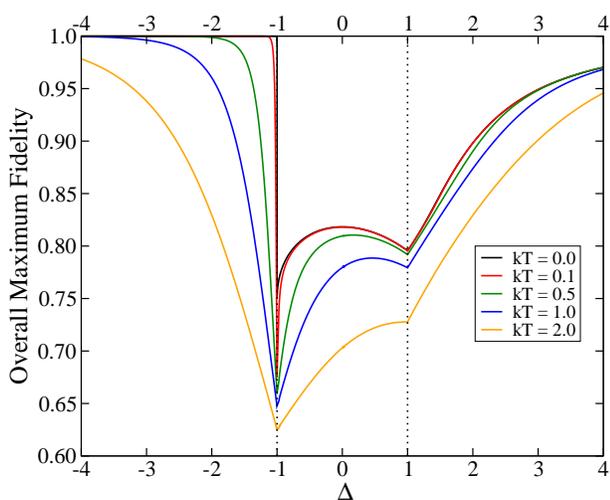}
\caption{\label{fig_max_geral}(color online) $\overline{\mathcal{F}}$, Eq.~(\ref{fmax}), as a function of the anisotropy 
$\Delta$ that characterizes the XXZ model [Eq.~(\ref{Hxxz})]. 
Both QCPs are detected at $T=0$ and $T>0$ by a discontinuity in
the first derivative of $\overline{\mathcal{F}}$ with respect to $\Delta$.
In the curves above, temperature increases from top to bottom.
Here and in all other figures all quantities are dimensionless.}
\end{figure}

In Fig. \ref{fig_max_geral} we plot $\overline{\mathcal{F}}$ as a function of
$\Delta$ for several values of $T$. It is clear from 
Fig. \ref{fig_max_geral} that $\overline{\mathcal{F}}$ detects both QCPs when
$T=0$ and for $T>0$. It is worth mentioning that for $kT \gtrsim 0.1$ 
the EoF
is already zero before, at, and after the QCP $\Delta = -1$ 
\cite{wer10b}.
The cusp-like behavior of $\overline{\mathcal{F}}$ 
at the two QCPs is similar to that observed for the 
TQD \cite{wer10b}. The discontinuity of the first derivative 
of $\overline{\mathcal{F}}$ with respect to the tuning parameter $\Delta$ is 
related to the fact that at the two QCPs the roles of 
$|\langle\sigma_2^z\sigma_{3}^z\rangle|$ and $|\langle\sigma_2^x\sigma_{3}^x\rangle|$ 
are exchanged. For instance, in one phase the maximum of $\overline{\mathcal{F}}$
is a function of $|\langle\sigma_2^z\sigma_{3}^z\rangle|$ while at the other phase
it is a function of $|\langle\sigma_2^x\sigma_{3}^x\rangle|$.

If we now employ Eq.~(\ref{Flangle}), maximized over the
sets $S_k$ of unitary operations available to Bob, we get (see Appendix
\ref{apB})
\begin{eqnarray}
\langle\overline{\mathcal{F}}\rangle \hspace{-.1cm}&\hspace{-.1cm}=\hspace{-.1cm}& 
\hspace{-.05cm}\max_{\{S_k\}}{\hspace{-.05cm}\langle\overline{F}(S_k)\rangle}
\nonumber \\
\hspace{-.1cm}&\hspace{-.1cm}=\hspace{-.1cm}&\hspace{-.05cm}\max{\hspace{-.1cm}\left[\frac{3 \!+\! 2 |\langle\sigma_2^x\sigma_{3}^x\rangle|\!-\!
\langle\sigma_2^z\sigma_{3}^z\rangle}{6}, 
\frac{3 \!+\! \langle\sigma_2^z\sigma_{3}^z\rangle}{6}\right]}\hspace{-.05cm}.
\label{fmed}
\end{eqnarray}

\begin{figure}[!ht]
\includegraphics[width=8cm]{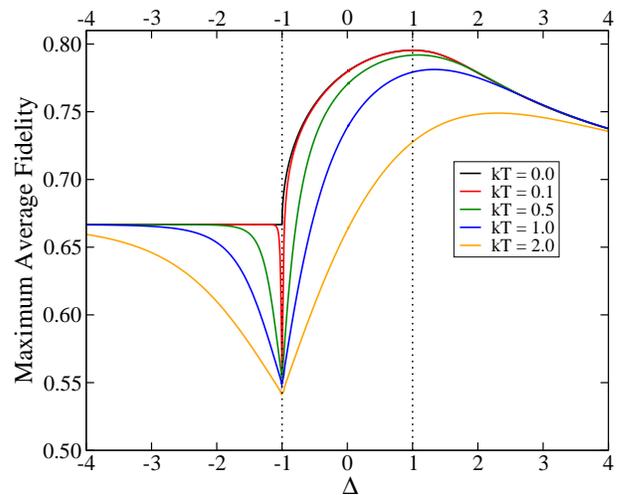}
\caption{\label{fig_med_geral}(color online) Same as Fig. \ref{fig_max_geral}
but now we plot  
$\langle\overline{\mathcal{F}}\rangle$, Eq.~(\ref{fmed}). The QCP at $\Delta=-1$ 
is detected at $T=0$ and $T>0$ by a discontinuity in
the first derivative of $\langle\overline{\mathcal{F}}\rangle$ with respect to 
$\Delta$. The other QCP is obtained at $T=0$ noting that
$\langle\overline{\mathcal{F}}\rangle$ is maximal at $\Delta = 1$. For finite $T$,
the maximum is displaced to greater values of $\Delta$. For $kT \lesssim 1.0$ these
maxima lie close together and by extrapolating to $kT \rightarrow 0$ we can infer this QCP by working with finite $T$ data. In the curves above, temperature increases from top to bottom.}
\end{figure}

Looking at Fig. \ref{fig_med_geral} we realize that the QCP located at 
$\Delta = -1$,
associated with a first order QPT, is clearly detected for $T=0$ and $T>0$. The 
cusp-like behavior of $\langle\overline{\mathcal{F}}\rangle$ at $\Delta=-1$
for $T=0$ and $T>0$ clearly indicates a QPT.
The other QCP, $\Delta = 1$, related to a continuous QPT is detected at $T=0$ by
noting that $\langle\overline{\mathcal{F}}\rangle$ has its global maximum exactly 
at $\Delta = 1$. For finite $T$ this maximum is displaced to higher
values of $\Delta$. Contrary to $\overline{\mathcal{F}}$,  
$\langle\overline{\mathcal{F}}\rangle$ does not have a
cusp at $\Delta=1$. However, if we work with 
both the maximum and minimum of $\langle\overline{F}(S_k)\rangle$, 
we can get a cusp-like behavior at both QCPs (see Appendix \ref{apC}).

The usefulness of $\overline{\mathcal{F}}$ and 
$\langle\overline{\mathcal{F}}\rangle$ to pinpoint a
QCP is not restricted to the XXZ model. They are as good as TQD \cite{wer10b} to 
detect for $T>0$ the QCP of the XXX model (see Appendix \ref{apD}). It remains an
open problem to check if the ideas here presented apply to the detection of
pseudo-transitions \cite{tim11,sou20}.

\section{Discussion and conclusion} 

As already highlighted above, the operational
and experimental interpretation of the fidelities is very 
clear and straightforward. For the XXZ and XXX model with no fields, 
whose two-qubit density matrix is given by Eq.~(\ref{rhoAB}), 
the experimental demands to implement the present 
proposal are further reduced. Since for those models 
$\overline{F}(|\psi\rangle,S_k)= F_j(| \psi \rangle,S_k)$, 
we do not need to implement any unitary correction $U_j$ on the teleported 
qubit to obtain the fidelities. We just need to 
separate the data into four sets, each one corresponding to the four possible 
outcomes of Alice's BMs. In this way, we automatically get
the mean fidelities $\overline{F}(|\psi\rangle,S_k)$ 
related to each one of the four sets $S_k$.  
Teleporting a representative sample of qubits covering the Bloch sphere, 
we obtain $\overline{\mathcal{F}}$ 
picking from this sample of teleported 
states the case yielding the greatest fidelity and, averaging over all cases, 
we get $\langle\overline{\mathcal{F}}\rangle$.
To fully execute
the teleportation protocol we also 
need to be able to implement the BMs on qubits 1 and 2. 
The BMs can be made by applying a controlled-not 
(CNOT) operation on those qubits \cite{ron15,bra19,xie19,noi22,xue22,mad22,xie22} 
followed by a Hadamard operation on the control qubit and a measurement of those spins in the computational basis \cite{nie00}. 
For instance, if after the previous prescription we see spins 1 and 2 
pointing up ($|00\rangle$) or down ($|11\rangle$), 
it means that we have projected them onto
the Bell state $|\Phi^+\rangle$ or $|\Psi^-\rangle$ \cite{nie00}.  

From the theoretical point of view, and similarly to EoF and QD, we
need the two-qubit density matrix, Eq.~(\ref{rhoAB}), to compute the fidelity. 
In a more general scenario (higher spins), 
we need the bipartite 
density matrix describing two $N$-dimensional systems. 
However, and contrary to EoF and QD, the computational
resources needed to compute the fidelity 
are less demanding. To compute the maximum average fidelity,
Eq.~(\ref{fmed}), we just need to repeat for each one of the four sets of unitary
operations $S_k$ the calculation of the 
average fidelity, Eq.~(\ref{Flangle}). 
The calculation of the latter is a very
simple matter and can be efficiently 
scaled to an $N$-dimensional input state $|\psi \rangle$ \cite{gor06}.
To calculate the overall maximum fidelity, Eq.~(\ref{fmax}), we have
to repeat the maximization of Eq.~(\ref{Fbar}) over all input states $|\psi \rangle$ four times (for each one of the four sets of unitary operations $S_k$). 
Since an $N$-dimensional pure state is describe by $2N-2$ independent parameters \cite{footnote2}, we will face an optimization problem involving $2N-2$ free 
variables. For high values of $N$ this is not a simple problem but it is less demanding
than solving the corresponding optimization problem to determine the QD, where
we must minimize the conditional entropy over all sets of 
generalized measurements (POVMs) \cite{oll01,hen01,hua14}. These POVMs are
$N\times N$ matrices and the number of free parameters increases
faster than linearly with $N$ \cite{hua14}. The intuitive reason for this 
difference in computational demand rests on the fact that for the overall
maximum fidelity we optimize over a single pure state while for QD the 
optimization problem is equivalent to the complexity of determining 
the EoF, 
whose optimization is done over all ensembles of pure states 
into which $\rho_{23}$ can be decomposed \cite{hua14}.

Summing up, we have presented two teleportation based theoretical tools to detect QCPs 
at finite $T$ equivalent to TQD, the most reliable QCP detector 
for finite $T$ known to date. 
Both tools work without the knowledge of the order parameter associated with 
the QPT. The tools here presented have two features that set them apart from 
TQD and other quantum information theory based QCP detectors. First, they have a straightforward experimental interpretation and can in principle be directly 
measured in the laboratory. 
Second, from a theoretical point of view we need fewer computational resources 
to calculate them when compared to TQD, with one of these tools,
the average fidelity, 
easily scalable to an $N$-dimensional spin system.

\begin{acknowledgments}
GR thanks the Brazilian agency CNPq
(National Council for Scientific and Technological Development) for funding and 
CNPq/FAPERJ (State of Rio de Janeiro Research Foundation) for financial support
through the National Institute of Science and Technology for Quantum Information.
\end{acknowledgments}

\appendix

\section{Two-point correlation functions}
\label{apA}

The XXZ model we studied is given by Eq.~(1) 
of the main text. In thermal equilibrium with a reservoir at temperature
$T$, the non-null two-point correlation functions are 
$\med{\sigma_j^x\sigma_{j+1}^x}=\med{\sigma_j^y\sigma_{j+1}^y}$ and 
$\med{\sigma_j^z\sigma_{j+1}^z}$. 
The techniques to solve this problem in the thermodynamic limit (infinite chain) 
were developed in Refs. \cite{klu92,bor05,boo08,tri10} and they were carefully reviewed, adapted, and implemented for the present context in Ref. \cite{wer10b}. 

\begin{figure}[!ht]
\includegraphics[width=8cm]{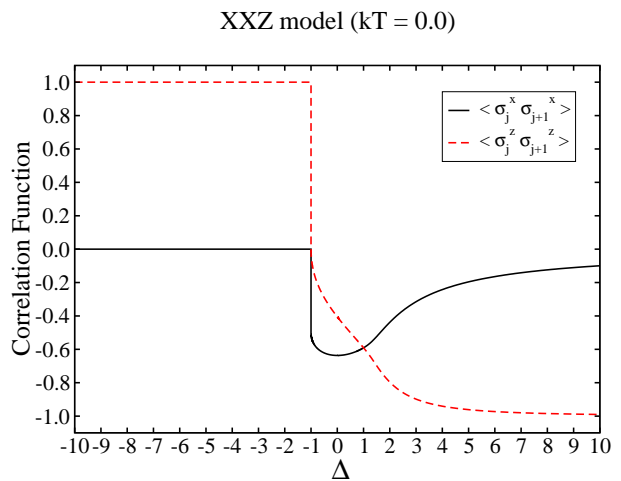}
\caption{\label{fig1_sup}(color online)  Two-point correlation functions 
in the thermodynamic limit at $T=0$
as a function of the tuning parameter $\Delta$.}
\end{figure}

\begin{figure}[!ht]
\includegraphics[width=8cm]{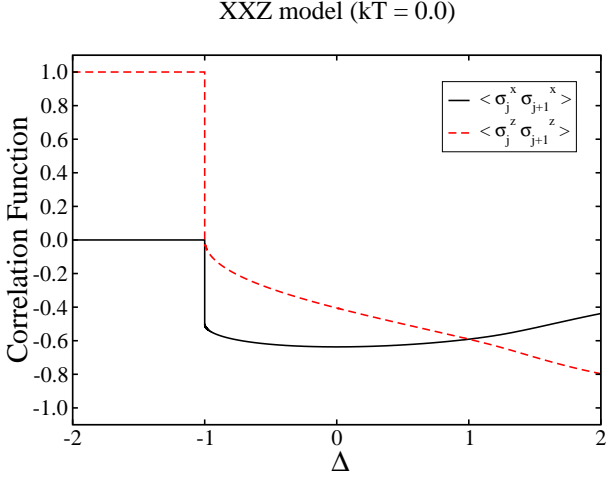}
\caption{\label{fig1zoom_sup}(color online)  Two-point correlation functions 
in the thermodynamic limit at $T=0$
as a function of the tuning parameter $\Delta$.}
\end{figure}

At the absolute zero temperature, the two-point correlation functions are given 
by Figs. \ref{fig1_sup} and \ref{fig1zoom_sup} \cite{wer10b}.

\begin{figure}[!ht]
\includegraphics[width=8cm]{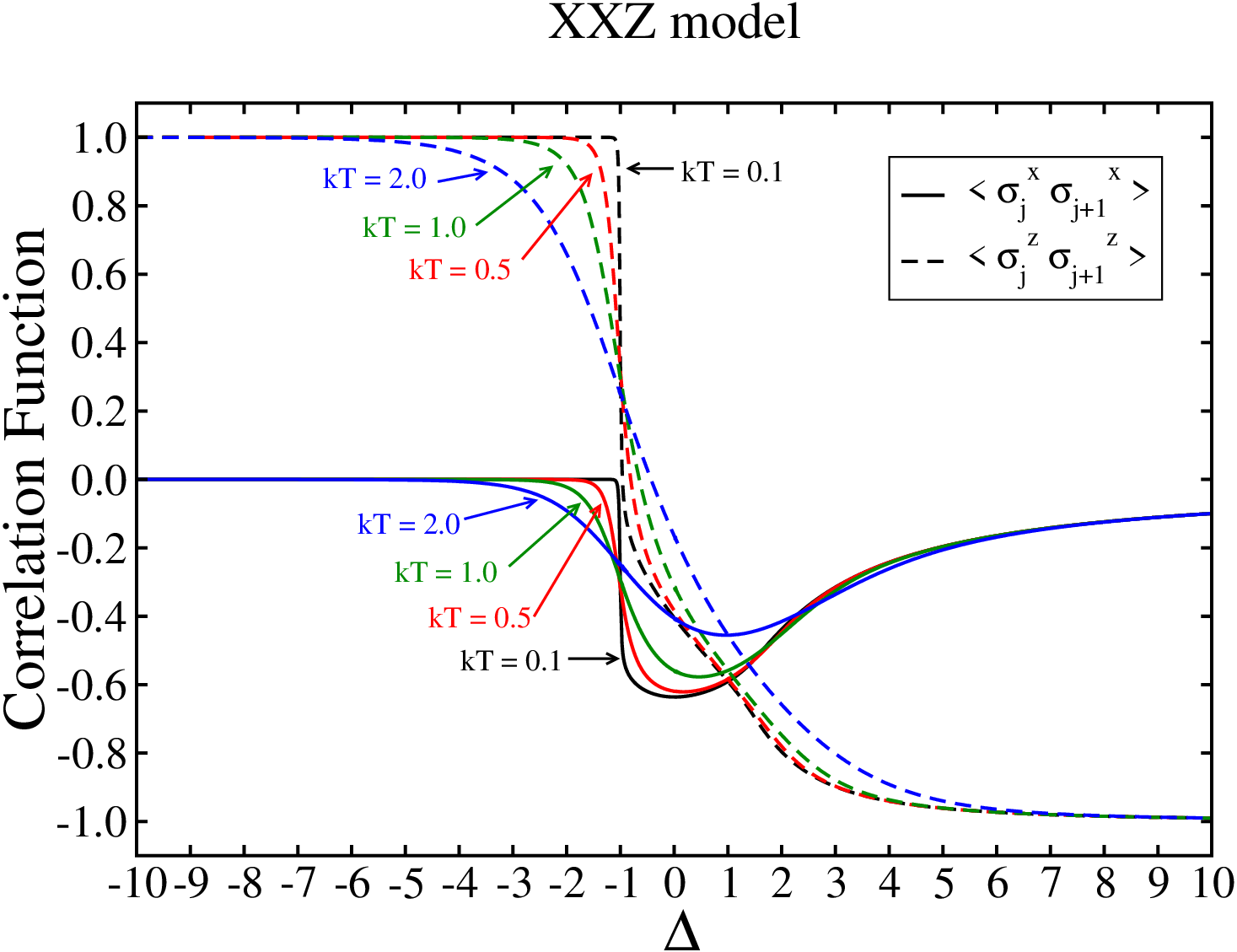}
\caption{\label{fig2_sup}(color online)  Two-point correlation functions 
in the thermodynamic limit for $T>0$
as a function of the tuning parameter $\Delta$.}
\end{figure}

\begin{figure}[!ht]
\includegraphics[width=8cm]{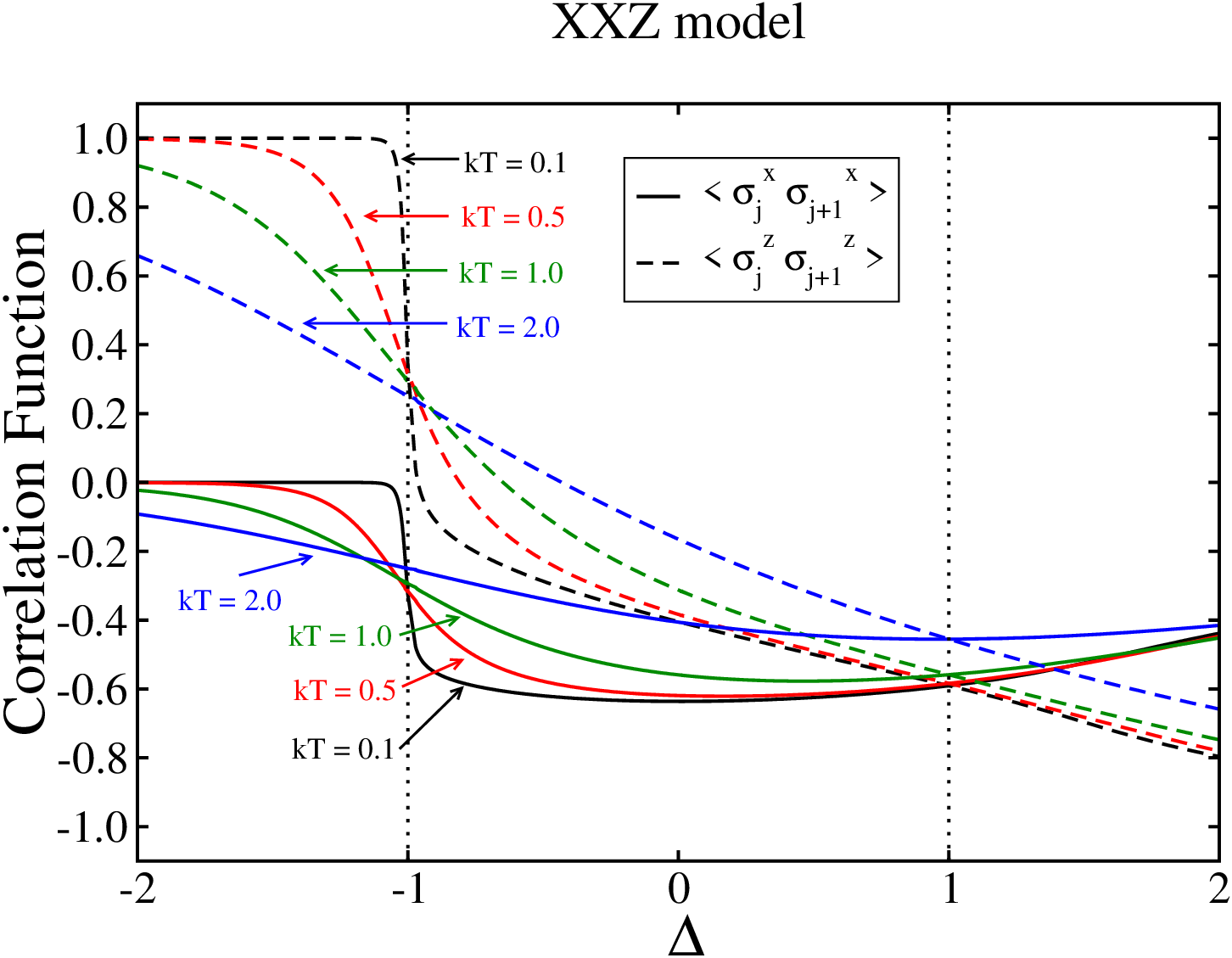}
\caption{\label{fig2zoom_sup}(color online)  Two-point correlation functions 
in the thermodynamic limit for $T>0$
as a function of the tuning parameter $\Delta$.}
\end{figure}

For finite $T$, the behavior of the two-point correlation functions is given by
Figs. \ref{fig2_sup} and \ref{fig2zoom_sup} \cite{wer10b}.

Note that at the quantum critical points (QCPs) the two-point 
correlation function having the greatest magnitude changes. 
This is particularly clear for low values of 
$kT$. Before $\Delta=-1$, $|\med{\sigma_j^z\sigma_{j+1}^z}| > 
|\med{\sigma_j^x\sigma_{j+1}^x}|$ while after $\Delta=-1$, 
$|\med{\sigma_j^x\sigma_{j+1}^x}| > |\med{\sigma_j^z\sigma_{j+1}^z}|$.
This behavior is also seen before and after the other QCP at $\Delta = 1$ and 
it is the reason for the cusp like behavior of the 
overall maximum fidelity $\overline{\mathcal{F}}$ at both QCPs (see main text).

\section{Obtaining $\overline{\mathcal{F}}$ and 
$\langle \overline{\mathcal{F}} \rangle$}
\label{apB}

In the XXZ model, 
we have $Q_j(|\psi\rangle)=1/4$ and
$F_j(| \psi \rangle,S_k)=F_{j'}(| \psi \rangle,S_k)$ 
for any $j,j'=\Psi^\mp, \Phi^\mp$.
Thus, a direct calculation 
using Eqs.~(3)-(5) and (11) 
of the main text allows us to write Eq.~(12) 
of the main text as follows
\begin{eqnarray}
\overline{F}(| \psi \rangle,S_{\Psi^-}) 
&=& f(r,-xx,zz), \\ 
\overline{F}(| \psi \rangle,S_{\Psi^+}) 
&=& f(r,xx,zz), \\ 
\overline{F}(| \psi \rangle,S_{\Phi^-}) 
&=& g(r,\gamma,-xx,zz), \\
\overline{F}(| \psi \rangle,S_{\Phi^+}) 
&=& g(r,\gamma,xx,zz), 
\end{eqnarray}
where
\begin{eqnarray}
f(r,xx,zz)\!&\!=\!&\! [1 \!+ 4 r^2 (1 - r^2) (xx + zz ) \!-\! zz]/2, \\
g(r,\gamma,xx,zz)&=&[1 + (1 - 2 r^2)^2 zz \nonumber \\
&&+ 4 r^2 (1 - r^2) xx \cos(2 \gamma)]/2, \\
xx &=& \med{\sigma_j^x\sigma_{j+1}^x}, \\
zz &=& \med{\sigma_j^z\sigma_{j+1}^z}.
\end{eqnarray}

Computing the extrema of $f(r,xx,zz)$, i.e., solving
\begin{equation}
\frac{\partial f}{\partial r} = 0,
\end{equation}
we immediately get for $r\geq 0$
\begin{equation}
r=0, 1/\sqrt{2}. 
\end{equation}
This means, according to Eq.~(3) 
of the main text, that the states leading to the extrema of 
$\overline{F}(| \psi \rangle,S_{\Psi^\mp})$ are, up to an overall phase, 
either $|1\rangle$ or $(|0\rangle + e^{i\gamma}|1\rangle)/\sqrt{2}$.

If we now compute the extrema of $g(r,\gamma,xx,zz)$, namely, if we solve
\begin{eqnarray}
\frac{\partial g}{\partial r} &=& 0, \\
\frac{\partial g}{\partial \gamma} &=& 0,
\end{eqnarray}
we obtain 
\begin{eqnarray}
(r;\gamma)&=&(0; 0\leq \gamma < 2\pi), \\
(r;\gamma)&=&(1/\sqrt{2}; 0, \pi/2, \pi, 3\pi/2).
\end{eqnarray}
This implies that the input states leading to the extrema of 
$\overline{F}(| \psi \rangle,S_{\Phi^\mp})$ are, up to an overall phase, 
$|1\rangle$, $(|0\rangle \pm |1\rangle)/\sqrt{2}$, and  
$(|0\rangle \pm i |1\rangle)/\sqrt{2}$.

Inserting the corresponding values of $r$ and $\gamma$ 
for the extrema of the fidelity we get
\begin{eqnarray}
\overline{F}(r=0,S_{\Psi^\mp}) &=& (1-zz)/2, \label{fpsi}\\
\overline{F}(r=1/\sqrt{2},S_{\Psi^\mp}) &=& (1\mp xx)/2, \\
\overline{F}(r=0,S_{\Phi^\mp}) &=& (1+zz)/2, \\
\overline{F}(r=1/\sqrt{2},\gamma,S_{\Phi^\mp}) &=& [1\mp xx\cos(2\gamma)]/2.
\label{fphi}
\end{eqnarray}
Therefore, Eqs.~(\ref{fpsi})-(\ref{fphi}) lead to
\begin{equation}
\overline{\mathcal{F}}= \max_{\{|\psi\rangle,S_k\}}{\overline{F}(|\psi\rangle,S_k)}
=\max{\left[\frac{1 + |zz|}{2}, \frac{1 + |xx|}{2}\right]},
\end{equation}
which is exactly Eq.~(14) 
of the main text.

Moving to the calculation of $\langle \overline{\mathcal{F}} \rangle$, we first note that if we
write the input state as 
\begin{equation}
|\psi \rangle = a |0\rangle + b |1\rangle,
\end{equation}
where $|a|^2+|b|^2=1$,
Eq.~(13) 
of the main text becomes
\begin{eqnarray}
\langle \overline{F}(S_{\Psi^\mp}) \rangle &=&  
[(\langle|a|^4\rangle +  \langle |b|^4\rangle) (1 - zz) \nonumber \\
&&+ 2 \langle|a b|^2\rangle (1 \mp 2 xx + zz)]/2. \label{fpsi-}
\end{eqnarray}

If $|a|^2$, the probability of finding the input in the state $|0\rangle$, 
and the relative phase between the complex numbers $a$ and $b$ are given by two 
independent continuous uniform distributions, we have \cite{gor06,rig15}
\begin{equation}
\langle|a|^4\rangle = \langle|b|^4\rangle = 1/3, \hspace{.5cm} 
\langle|a b|^2\rangle = 1/6. \label{averages1}
\end{equation}

Note that we will obtain the same averages if we use the Bloch sphere representation
for the input state \cite{nie00},
\begin{equation}
|\psi \rangle = \cos(\theta/2) |0\rangle + \sin(\theta/2)e^{i\varphi} |1\rangle,
\end{equation}
and average over the whole Bloch sphere. In this case, using the notation
of Eq. (13) of the main text, 
$d|\psi\rangle = dA = \sin\theta d\theta d\varphi$ is the element of area of a 
unit sphere written in spherical polar coordinates, 
$\mathcal{P}(|\psi\rangle)=1/(4\pi)$, $0\leq \theta \leq \pi$, and 
$0\leq \varphi \leq 2\pi$.

Thus, using Eq.~(\ref{averages1}) we get for Eq.~(\ref{fpsi-}),
\begin{equation}
\langle \overline{F}(S_{\Psi^\mp}) \rangle = (3 \mp 2 xx - zz)/6. \label{fpsi+-}
\end{equation}
In an analogous way we obtain
\begin{equation}
\langle \overline{F}(S_{\Phi^\mp}) \rangle = (3 + zz)/6, \label{fphi+-}
\end{equation}
where to arrive at Eq.~(\ref{fphi+-}) we also used that
\begin{equation}
\langle (a^* b)^2\rangle = \langle(a b^*)^2\rangle = 0, 
\end{equation}
with $a^*(b^*)$ denoting the complex conjugate of $a(b)$.
Finally, looking at Eqs.~(\ref{fpsi+-}) and (\ref{fphi+-}) we get
\begin{eqnarray}
\langle\overline{\mathcal{F}}\rangle &=& 
\max_{\{S_k\}}{\langle\overline{F}(S_k)\rangle}
\nonumber \\
&=&\max{\left[\frac{3 + 2 |xx|-
zz}{6}, 
\frac{3 + zz}{6}\right]},
\end{eqnarray}
which is Eq.~(15) 
of the main text.

\section{Looking deeper at $\overline{\mathcal{F}}$ and 
$\langle \overline{\mathcal{F}} \rangle$}
\label{apC}

In order to better understand the behavior of the curves shown in Fig. 2 
of the main 
text, we will study the behavior of the following quantity,
\begin{equation}
\overline{F}(S_k)= \max_{\{|\psi\rangle\}}{\overline{F}(|\psi\rangle,S_k)}. 
\label{fsk}
\end{equation}
Equation (\ref{fsk}) is the mean fidelity, Eq.~(12) 
of the main text, maximized over the input states only. We want to investigate 
the behavior of $\overline{F}(S_k)$ for each one of the four possible values of $k$,
\begin{eqnarray}
\overline{F}(S_{\Psi^\mp}) &=& \max{\left[\frac{1 - zz}{2}, \frac{1 \mp xx}{2}\right]},
\label{f1max}\\
\overline{F}(S_{\Phi^\mp}) &=& \max{\left[\frac{1 + zz}{2}, \frac{1 + |xx|}{2}\right]}.
\label{f4max}
\end{eqnarray}
Proceeding in this way we will be able to trace back 
which expression is responsible to 
the behavior of $\overline{\mathcal{F}}$ shown in Fig. 2 
of the main text.

\begin{figure}[!ht]
\includegraphics[width=8cm]{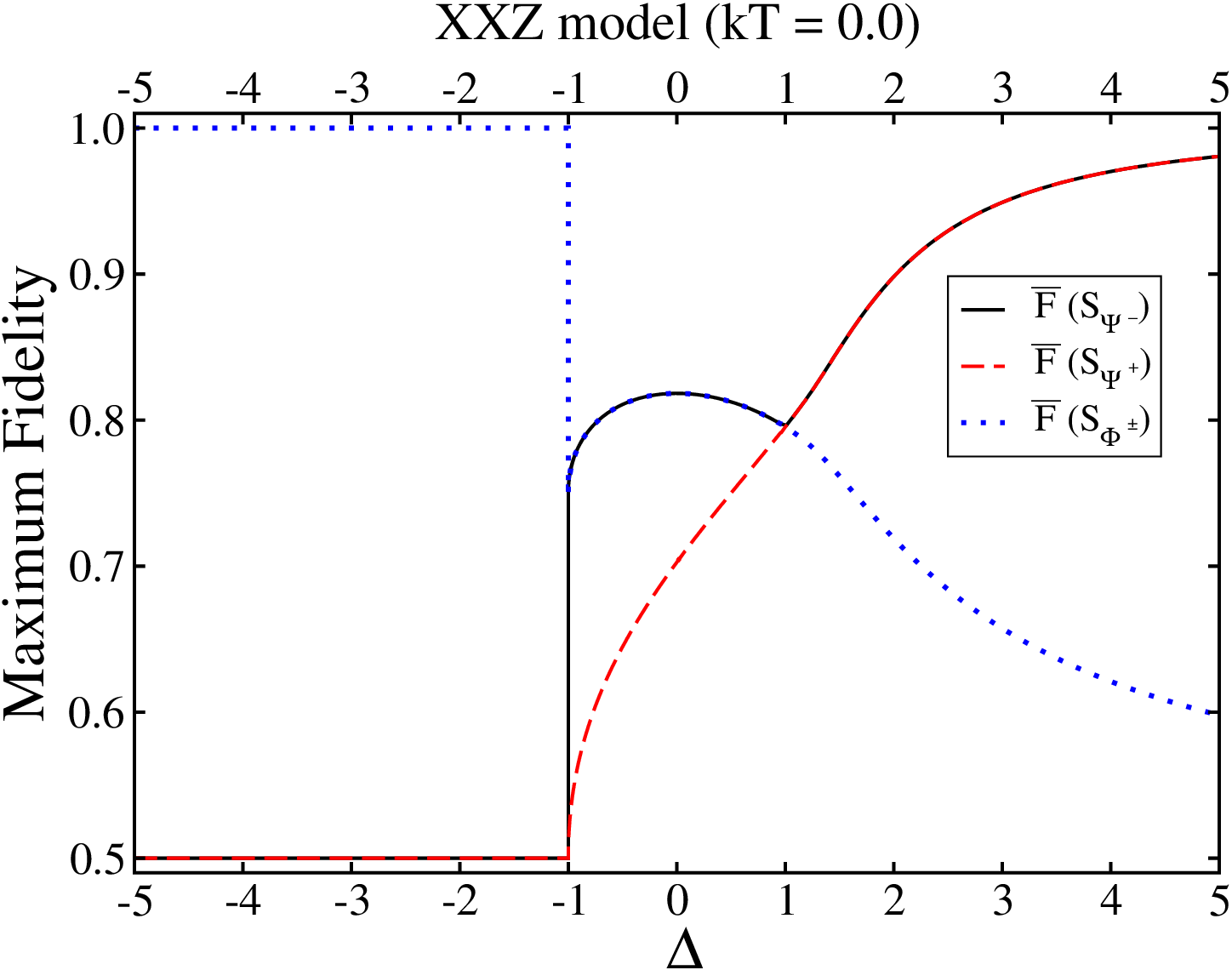}
\caption{\label{fig_max1a4_kT0}(color online) $\overline{F}(S_{k})$ as a 
function of $\Delta$ at the absolute zero temperature.}
\end{figure}

\begin{figure}[!ht]
\includegraphics[width=8cm]{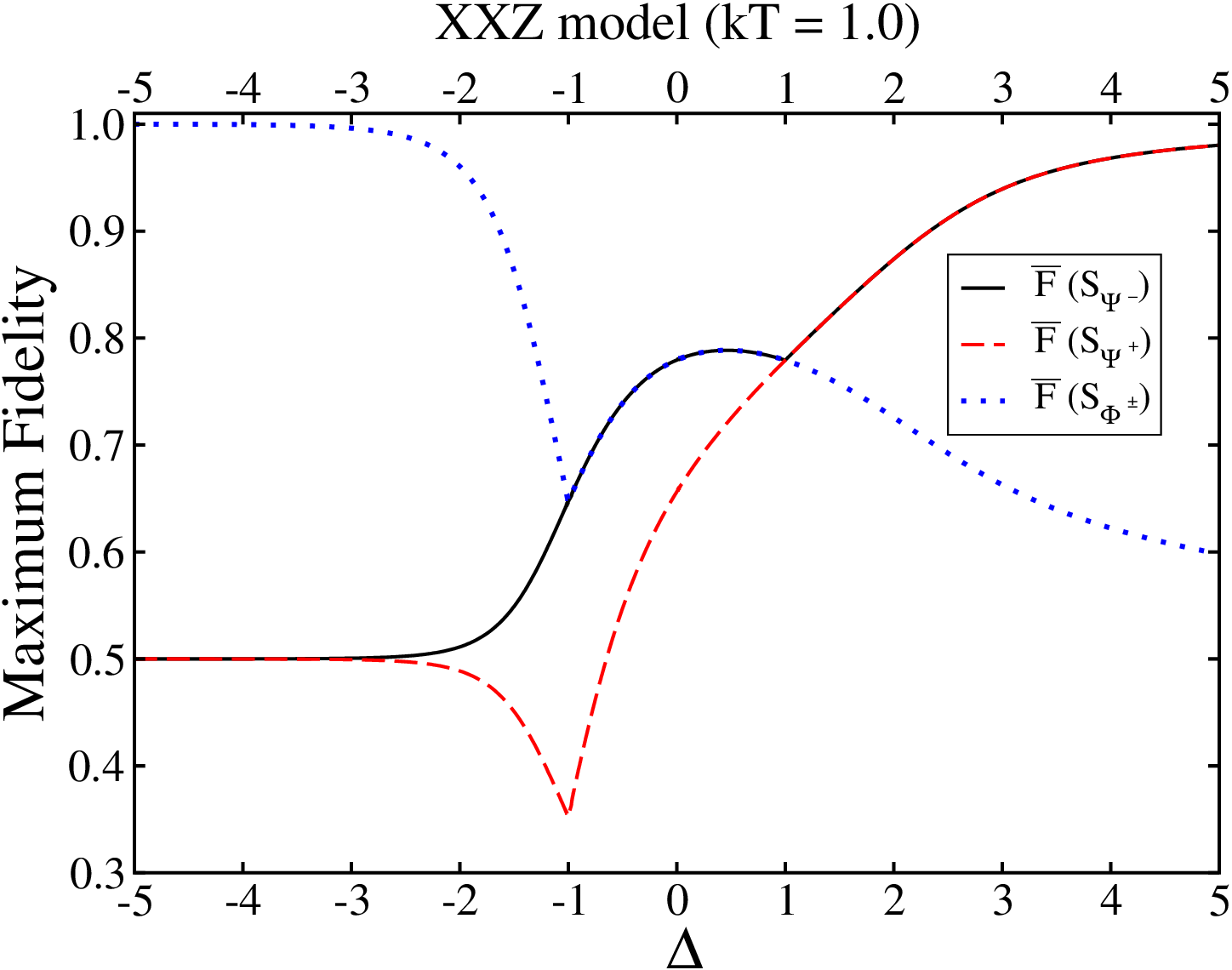}
\caption{\label{fig_max1a4_kT1p0}(color online) Same as Fig. \ref{fig_max1a4_kT0}
but at $kT=1.0$.}
\end{figure}

In Fig. \ref{fig_max1a4_kT0} we show Eqs.~(\ref{f1max}) and (\ref{f4max}) at $T=0$
and in Fig. \ref{fig_max1a4_kT1p0} at $T>0$ as a function of the tuning parameter
$\Delta$.

Looking at both Figs. \ref{fig_max1a4_kT0} and \ref{fig_max1a4_kT1p0}, we realize
that before the first QCP, $\Delta = -1$, the overall maximum fidelity  
$\overline{\mathcal{F}}$ is given by 
$\overline{F}(S_{\Phi^\mp})$. Between
the two QCPs we have either $\overline{F}(S_{\Phi^\mp})$ or 
$\overline{F}(S_{\Psi^-})$ as the maximum fidelity. And after the second 
QCP, $\Delta = 1$, it is $\overline{F}(S_{\Psi^\mp})$ that dominates.
The change of which $\overline{F}(S_k)$ dominates at the QCPs 
is the reason for the
cusp-like behavior of $\overline{\mathcal{F}}$.

A similar analysis allows us to understand the behavior of the maximum average 
fidelity $\langle \overline{\mathcal{F}} \rangle$ as shown in Fig. 3 
of the main text. We now investigate $\langle\overline{F}(S_k)\rangle$, Eq.~(13)
of the main text, for each one of the four possible values of $k$,
\begin{eqnarray}
\langle\overline{F}(S_{\Psi^\mp})\rangle &=& (3 \mp 2 xx-zz)/6, \\
\langle\overline{F}(S_{\Phi^\mp})\rangle &=& (3 + zz)/6.
\end{eqnarray}

\begin{figure}[!ht]
\includegraphics[width=8cm]{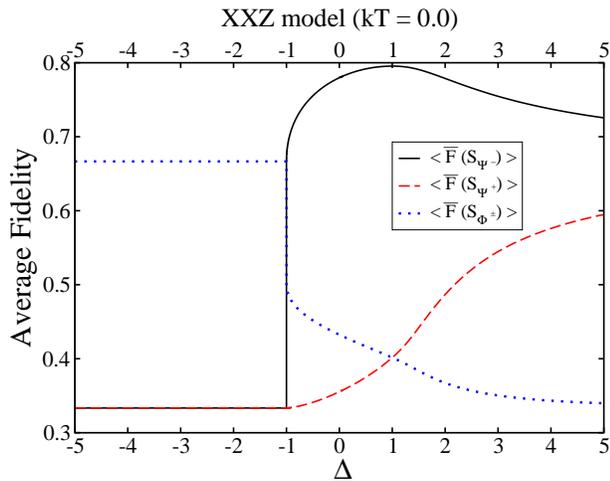}
\caption{\label{fig_med1a4_kT0}(color online) $\langle \overline{F}(S_{k})\rangle$ 
as a function of $\Delta$ at the absolute zero temperature.}
\end{figure}
\begin{figure}[!ht]
\includegraphics[width=8cm]{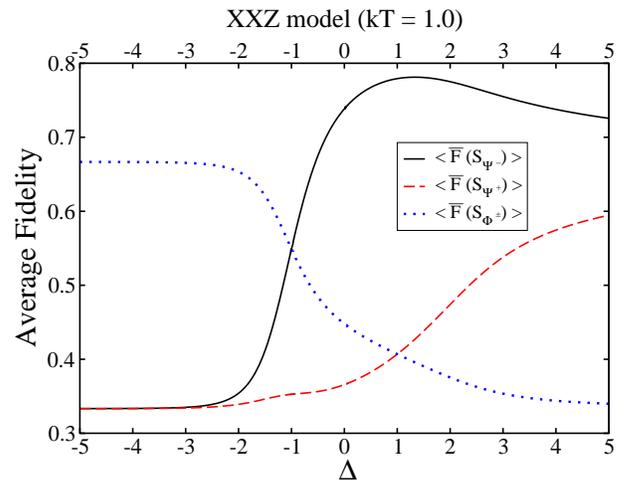}
\caption{\label{fig_med1a4_kT1p0}(color online) Same as Fig. \ref{fig_med1a4_kT0}
but at $kT=1.0$.}
\end{figure}

Looking at Figs. \ref{fig_med1a4_kT0} and \ref{fig_med1a4_kT1p0}, we note
that before the first QCP, $\Delta = -1$, the maximum average fidelity  
$\langle\overline{\mathcal{F}}\rangle$ is given by 
$\langle\overline{F}(S_{\Phi^\mp})\rangle$. After the first QCP we have 
$\langle\overline{F}(S_{\Psi^-})\rangle$ as the maximum average fidelity. 
This trend continues even after the second QCP, where 
$\langle\overline{F}(S_{\Psi^-})\rangle$ achieves its maximum. That is why 
we do not see any cusp-like behavior at $\Delta=1$, the second QCP. At this 
QCP, there is no change in the function maximizing the average fidelity. Before,
at, and after $\Delta=1$, it is 
always $\langle\overline{F}(S_{\Psi^-})\rangle$ that
maximizes the average fidelity.

Incidentally, looking at Figs. \ref{fig_med1a4_kT0} and \ref{fig_med1a4_kT1p0}, we 
realize that by monitoring both the maximum and the minimum average fidelity we 
can detect both QCPs via a cusp-like behavior. Indeed, monitoring the maximum
average fidelity we see a cusp-like behavior at $\Delta=-1$, the first QCP. This
is what is shown in Fig. 3 
of the main text. However, if we plot the minimum average fidelity,
\begin{eqnarray}
\langle\overline{\mathcal{F}}\rangle_{min} &=& 
\min_{\{S_k\}}{\langle\overline{F}(S_k)\rangle},
\end{eqnarray}
we will see a cusp-like behavior not in the first but in the second QCP. At 
$\Delta = 1$ we see that the function giving the minimum average fidelity 
changes. Before $\Delta = 1$ the minimum is achieved by 
$\langle\overline{F}(S_{\Psi^+})\rangle$ and after it the minimum is given by
$\langle\overline{F}(S_{\Phi^\mp})\rangle$. This change of the function minimizing
the average fidelity leads to a cusp-like behavior at this QCP. 
As such, by monitoring both the maximum and minimum of 
$\langle\overline{F}(S_k)\rangle$, we can build an input state independent 
fidelity as sharp as the state dependent fidelity in 
detecting both QCPs, with the advantage that the computation of the 
former quantity is easily scalable to high spin systems.

\begin{figure}[!ht]
\includegraphics[width=8cm]{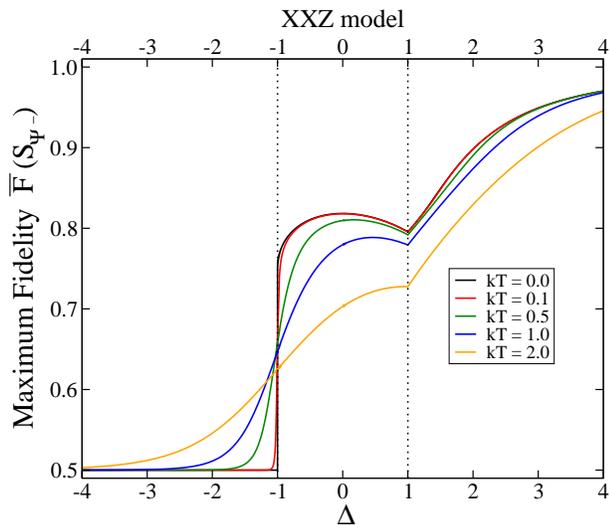}
\caption{\label{fig_max1}(color online) $\overline{F}(S_{\Psi^-})$ 
as a function of $\Delta$ for several values of temperature. Temperatures 
increase from top to bottom when $\Delta > -1$.}
\end{figure}

\begin{figure}[!ht]
\includegraphics[width=8cm]{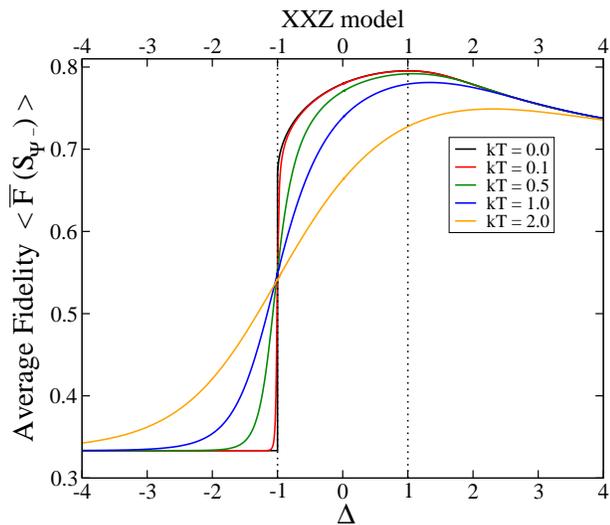}
\caption{\label{fig_med1}(color online) $\langle \overline{F}(S_{k})\rangle$ 
as a function of $\Delta$ for several values of temperature. Temperatures 
increase from top to bottom when $\Delta > -1$.}
\end{figure}

We finish this section showing $\overline{F}(S_{\Psi^-})$ and 
$\langle\overline{F}(S_{\Psi^-})\rangle$ (see Figs. \ref{fig_max1} and 
\ref{fig_med1}) for 
several values of $T$. Although these
quantities are not as good as $\overline{\mathcal{F}},
\langle\overline{\mathcal{F}}\rangle$, and
$\langle\overline{\mathcal{F}}\rangle_{min}$ to pinpoint a QCP, 
we can obtain a lot of information about where the QCPs are located
by working with them. We also obtain similar results working with 
$\overline{F}(S_k)$ and $\langle\overline{F}(S_k)\rangle$, 
where $k=\Psi^+,\Phi^\mp$. 


\section{The XXX model}
\label{apD}

The XXX model is given by the following Hamiltonian,
\begin{equation}
H = J \sum_{j=1}^{L}\left(
\sigma^{x}_{j}\sigma^{x}_{j+1} +
\sigma^{y}_{j}\sigma^{y}_{j+1} + 
\sigma^{z}_{j}\sigma^{z}_{j+1}\right). 
\label{Hxxx}
\end{equation}
This model is essentially the XXZ model with $\Delta = 1$, 
where we can now change the sign of the whole Hamiltonian by varying the 
parameter $J$. As before,  we use periodic boundary conditions. The XXX model
has one QCP, located at $J=0$. For $J<0$ we have the ferromagnetic phase and
for $J>0$ the antiferromagnetic one. 

\begin{figure}[!ht]
\includegraphics[width=8cm]{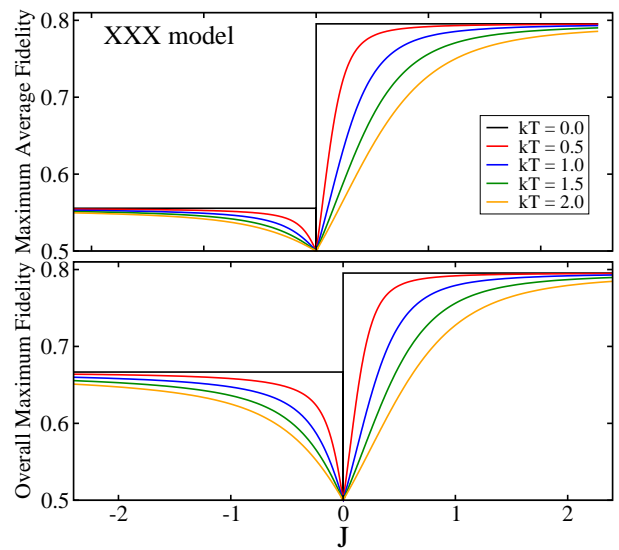}
\caption{\label{fig_med_max_geral}(color online) 
$\langle\overline{\mathcal{F}}\rangle$ 
(upper panel) and $\overline{\mathcal{F}}$  (lower panel) for a 
spin-1/2 chain in the thermodynamic limit. Temperatures 
increase from top to bottom.}
\end{figure}

In Fig. \ref{fig_med_max_geral} we show 
$\langle\overline{\mathcal{F}}\rangle$ 
and $\overline{\mathcal{F}}$ 
as a function of 
$J$ for several values of temperature. It is clear from those figures that 
both $\langle\overline{\mathcal{F}}\rangle$ and $\overline{\mathcal{F}}$ detect 
the QCP at finite $T$, similar to what one obtains computing the thermal
quantum discord \cite{wer10b}.
Also, for $kT \gtrsim 0.1$ the entanglement is already zero before, at, and after
the QCP \cite{wer10b}.


\begin{thebibliography}{200}
\bibitem{footnote1} TQD is the QD computed for systems
in equilibrium with a thermal reservoir at temperature $T$.
\bibitem{footnote1b} When we refer to Eq.~(\ref{Fbar}) we use the term 
``mean fidelity'' while we reserve the term ``average fidelity'' for 
Eq.~(\ref{Flangle}).
\bibitem{footnote2} We have $N$ complex coefficients leading to $2N$ real parameters. However, since an overall phase is irrelevant and we must satisfy the normalization constraint, we are left with $2N-2$ independent parameters.
\end{thebibliography}
\end{document}